\documentclass[12pt, twoside]{article}

\newcommand {\fract}[2]{\mbox{${\textstyle{\frac{#1}{#2}}}$}}

\begin{document}

\vspace*{-1cm}
\begin{flushright}
FTUV--01/1001\quad IFIC--00--67
\\
DAMTP--2000--136
\\
January 9, 2001
\\[1.5cm]
\end{flushright}

\begin{center}
\begin{Large}
\bfseries{Hidden supersymmetries in supersymmetric quantum mechanics}
\end{Large}

\vspace*{1.6cm}

\begin{large}
J.~A.~de~Azc\'arraga$^{a1}$, J.~M. Izquierdo$^{b1}$
and 
A.~J. Macfarlane$^{c}$\footnote{
j.a.de.azcarraga@ific.uv.es, izquierd@fta.uva.es,
a.j.macfarlane@damtp.cam.ac.uk}
\end{large}
\vspace*{0.6cm}

\begin{it}
$a$ Departamento de F\'{\i}sica Te\'orica, Universidad de Valencia
\\
and IFIC, Centro Mixto Universidad de Valencia--CSIC,
\\
E--46100 Burjassot (Valencia), Spain
\\[0.4cm]
$b$ Departamento de F\'{\i}sica Te\'orica, Universidad de Valladolid
\\
E--47011 Valladolid, Spain
\\[0.4cm]
$c$ Centre for Mathematical Sciences, D.A.M.T.P.
\\
Wilberforce Road, Cambridge CB3 0WA, UK\\
\end{it}

\end{center}
\vspace*{1cm}

\begin{abstract}
We discuss the appearance of additional, hidden supersymmetries 
for simple $0+1$ $Ad(G)$-invariant supersymmetric models and 
analyse some geometrical mechanisms that lead to them. It is shown 
that their existence depends crucially on the availability of odd order 
invariant skewsymmetric tensors on the (generic) compact Lie
algebra $\cal G$, and hence on the cohomology properties of the 
Lie algebra considered.
\end{abstract}

\newpage

\section{Introduction}

  In supersymmetric quantum mechanics
models with standard supersymmetry, the supercharges
$Q_a$ are related to the Hamiltonian $H$ via $\{Q_a,Q_b\}=H\delta_{ab}$,
$a,b=1,\dots, N$. In many of these models one can find additional 
or `hidden' supercharges $\tilde Q$ \cite{Gi.Ri.va, Ma.Mo.2}, 
involving 
the structure constants of a Lie algebra, and perhaps a Killing-Yano 
tensor \cite{Ki.Ya,Di.Ru}. The appearance of the Killing-Yano tensor in this 
context is not surprising, since it also plays a role in the existence of 
hidden symmetries \cite{Gi.Ru,Ta}. 

The additional supercharges are required to satisfy
\begin{equation}
  \{ Q_a,{\tilde Q}\}=0\quad ;                           \label{A1}
\end{equation}
hence $[{\tilde Q},H]=0$, so that the ${\tilde Q}$'s generate 
supersymmetries of the theory. We shall consider three models: 
one with bosonic superfields, and two with fermionic superfields 
(with $N=1$ and $N=2$ respectively) for which the bosonic component 
variables are auxiliary \cite{Macfarlane}. 

A typical example of the bosonic superfield case, in which the 
Lie algebra ${\cal G}$ is $su(2)$, is that of 
the non-relativistic motion of a spin-$\frac{1}{2}$ particle in the background 
field of a Dirac monopole \cite{De.Ma.Pe.va} (see also \cite{Ply}).
The case $G=SU(2)$ is, however, rather exceptional since it is the only group 
for which the structure constants of its algebra coincide with the fully 
antisymmetric tensor $(\epsilon_{ijk})$ of a 
$({\rm dim}\,G)$-dimensional space.
Thus, a natural question to ask is what generalisations are 
possible when simple (and compact) algebras 
$\cal G$ of rank $l>1$ are employed. Also, for $l>1$ there exist 
other available
skewsymmetric tensors (of odd order $>3$): they are provided by the 
higher order cocycles of the Lie algebra cohomology of $\cal G$. Thus,
in order to investigate the appearance of hidden charges in a group 
theoretical context, we look first at

\begin{equation}
{\tilde Q}_3={\dot x}_i f_{ij}\psi_j -\fract{1}{3}i f_{ijk}\psi_i\psi_j\psi_k
                                           \quad ,      \label{A2}
\end{equation}
where $x_i$ and $\psi_i$ are the position and fermionic coordinates 
respectively, $i=1,\dots,{\rm dim}\,{\cal G}$, and $f_{ij}$ is 
an antisymmetric 
second order Killing-Yano tensor \cite{Ki.Ya,Di.Ru}
associated with the structure constants $f_{ijk}$ in such a way that 
(\ref{A1}) holds, and second we look at

\begin{equation}
    {\tilde Q}_5= {\dot x}_i f_{ijkl}\psi_j\psi_k\psi_l -
 \fract{1}{5}i\Omega_{ijkpq}\psi_i\psi_j\psi_k\psi_p\psi_q   \label{A3}
\end{equation}
in various contexts. In (\ref{A3}), $f_{ijkl}=f_{[ijkl]}$ is a fourth order 
generalised Killing-Yano tensor and $\Omega_{ijkpq}$ is a fifth order totally 
antisymmetric invariant 
tensor, associated with the third order (Racah-Casimir) invariant
symmetric polynomial of such $\cal G$ 
as allow for one. This exists for $su(n)$, $n\geq 3$, and $su(3)$ will be 
good enough to illustrate the extent of most of our results  
when using the fifth order cocycle.

In the fermionic superfield case, it is possible to construct models 
where the only dynamical fields are fermionic, and whose Lagrangian 
includes an interaction term constructed using the structure constants 
of the simple Lie algebra. Our aim is to construct, in terms of the 
corresponding fermionic variables and the higher order cocycles, hidden 
supercharges in the $N=1$ and $N=2$ cases. When $N=1$  we shall restrict 
the Lie algebra to $su(n)$, whereas in the $N=2$ case the simple Lie 
algebra $\cal G$ will be unrestricted. The restriction reflects the
fact that the discussion of the $N=1$ case employs the identity
$C_{ijk}\Omega_{ijks_4\dots s_{2m-1}}=0$, which we believe holds 
for all $\cal G$, but for which explicit detailed proofs are 
available \cite{compilation} only for $su(n)$.

We consider only $Ad(G)$-invariant 
simple $0+1$ supersymmetry models.
In Sec. 2 we describe the case of a non-relativistic system
moving in a space that is the representation space of the 
adjoint representation of the symmetry group $G$, coupled to a background 
potential $A_i$. It is shown in sec. 3 that in the free case there exist 
non-standard supersymmetries associated with all the
 higher order cocycles. In the presence of the background
field $A_i$ (sec. 4), however, we find only one hidden supercharge for 
the lowest order cocycle {\it i.e.}, for 
that given by the structure constants of 
$\cal G$.

In Sec. 5, we consider a $N=1$ simple purely fermionic model
and show that one may also construct hidden 
supercharges from each of the $l$ $su(n)$ 
algebra cohomology cocycles. In 
Sec. 6 the case with $N=2$ is considered. It is shown that we may 
construct two hidden supercharges for each of the $l$ cocycles of the 
Lie algebra $\cal G$.

\section{A particle model with bosonic and fermionic degrees of freedom}

Let $G$ be a compact, simple Lie group of algebra $\cal G$, $[X_i,X_j]=
if_{ijk}X_k$. 
We set out from the superspace Lagrangian
\begin{equation}
    {\cal L}=\fract{1}{2}i{\dot \Phi}_iD \Phi_i+iq D\Phi_i A_i(\Phi)=
   K+\theta L   \ , \quad i=1,\dots,{\rm dim}\, {\cal G}\ ,             
\label{B1}
\end{equation}
where $\theta$ is a real Grassmann variable,
the $\Phi_i=\Phi_i(t,\theta)$  are scalar superfields, and
the covariant derivative $D$ and the generator of supersymmetry 
$Q$ are given by
\begin{equation}
\Phi_i(t,\theta) = x_i(t)+i\theta\psi_i(t)\ ,\quad
D = \partial_\theta-i\theta\partial_t\ ,\quad
Q = \partial_\theta+i\theta\partial_t    \ .               \label{B1A}
\end{equation}
The Lagrangian is invariant under 
the (real, adjoint) action of $G$. Using
\begin{eqnarray}
    D\Phi_i &=& i(\psi_i-\theta {\dot x}_i)\ ,\nonumber\\
   A_i(\Phi) &=& A_i(x)+i\theta\psi_j \partial_j A_i(x)\ ,  \label{B2}
\end{eqnarray}
the expansion (\ref{B1}) of $\cal L$ gives
\begin{eqnarray}
   K &=& -\fract{1}{2}{\dot x}_i\psi_i-q\psi_i A_i\label{B3}\ ,\\
  L &=& \fract{1}{2} {\dot x}_i{\dot x}_i+\fract{1}{2}i\psi_i{\dot\psi}_i
 +q{\dot x}_iA_i-\fract{1}{2} iqF_{ij}\psi_i\psi_j \ ,  \label{B4}
\end{eqnarray}
where 
\begin{equation}
F_{ij}=\partial_i A_j-\partial_j A_i \ .                  \label{efe}
\end{equation}
 We easily find the momenta and canonical commutators
\begin{equation}
   p_i=\frac{\partial L}{\partial {\dot x}_i}={\dot x}_i+qA_i\ ,\quad
  [x_i,p_j]=i\delta_{ij}\ ,\quad \{\psi_i,\psi_j\}=\delta_{ij}\ ,\label{B5}
\end{equation}
and compute
\begin{equation}
     H=\fract{1}{2} {\dot x}_i{\dot x}_i+\fract{1}{2} iqF_{ij}\psi_i\psi_j
    \ .                                                \label{B6}
\end{equation}
The standard supersymmetry that leaves the action for (\ref{B1}) 
invariant is $\delta\Phi_i=-i\epsilon Q\Phi_i$, {\it i.e.}
\begin{equation}
     \delta x_i=-i\epsilon\psi_i\ ,\quad \delta\psi_i=\epsilon{\dot x}_i \ .
                                                            \label{B7}
\end{equation}
Noether's theorem
\begin{equation}
    -i\epsilon Q= \delta x_i p_i+\delta\psi_i\frac{\partial L}{\partial{\dot 
\psi}_i}-i\epsilon K     \ ,                                  \label{B8}
\end{equation}
where the piece depending on $K$ comes from the quasi-invariance of the 
Lagrangian, gives for the conserved supercharge
\begin{equation}
     Q={\dot x}_i\psi_i\quad .                             \label{B9}
\end{equation}
It is easy to check that $Q$ above generates (\ref{B7}) by means of the 
canonical formalism, and that
\begin{equation}
Q^2=H                                                     \label{B10}
\end{equation}
reproduces the right hand side of (\ref{B6}).

\section{Hidden supersymmetries in the free case}

\subsection{The case of ${\tilde Q}_3$}

Here we shall put $q=0$ in expressions in Sec. 2. We intend first to 
seek an additional supersymmetry ${\tilde Q}_3$
such that $\{ Q,{\tilde Q}_3\}=0$ in the form
\begin{equation}
{\tilde Q}_3={\dot x}_i f_{ij}\psi_j-\fract{1}{3}if_{ijk}\psi_i\psi_j\psi_k
         \quad ,                                           \label{C1}
\end{equation}
where $f_{ij}$ is to be determined.
Since it is easier to work classically, we use the Dirac bracket 
formalism corresponding to (\ref{B5}), so that for $F$ and $G$ functions of 
dynamical variables
\begin{equation}
    \{ F,G\} = \frac{\partial F}{\partial x_l}\frac{\partial G}{\partial p_l}
 - \frac{\partial F}{\partial p_l}\frac{\partial G}{\partial x_l} 
 +i(-1)^F \frac{\partial F}{\partial \psi_l}\frac{\partial G}{\partial \psi_l}\ .
                                                           \label{C2}
\end{equation}
Using $Q=p_i\psi_i$ from (\ref{B9}) and (\ref{C1}), we get
\begin{equation}
     \{Q,{\tilde Q}_3 \}=-\psi_l{\dot x}_i f_{ij,l}\psi_j-
       i{\dot x}_l({\dot x}_i f_{il}-if_{ljk}\psi_j\psi_k) \ , \label{C3}
\end{equation}
where $f_{ij,l}=\frac{\partial}{\partial x^l}f_{ij}$.
The second term in the r.h.s. of (\ref{C3}) is zero if
\begin{equation}
   f_{ij}=-f_{ji}\ ,                                             \label{C4}
\end{equation}
and the other two cancel if
\begin{equation}
f_{i[j,l]}=f_{ijl}\ .                                              \label{C5}
\end{equation}
We could even have written $f_{ij,k}$ instead of $f_{ijk}$ in (\ref{C1}),
and then $f_{ij,k}$ may effectively be replaced by $f_{[ij,k]}$ in virtue of 
the $\psi_i\psi_j\psi_k$ factor. Then (\ref{C5}) is
\begin{equation}
    f_{i[j,l]}=f_{[ij,l]}    \quad ,                             \label{k}
\end{equation}
so that 
\begin{equation}
   \partial_{(l}f_{i)j}=0                                      \label{ky}
\end{equation}
or
\begin{equation}
     2f_{ij,l}=f_{li,j}+f_{jl,i}\ .                         \label{C6}
\end{equation}
Equations (\ref{k}) or (\ref{ky}) state that the derivative of the 
antisymmetric tensor (\ref{C4}) is also skewsymmetric, and characterise
$f_{ij}$ as Killing-Yano tensor \cite{Ki.Ya,Di.Ru}. One way to satisfy this 
condition sets
\begin{equation}
   f_{ij}=f_{ijk}x_k\ ,                                          \label{C7}
\end{equation}
giving 
\begin{equation}
   {\tilde Q}_3=L_i\psi_i+\fract{2}{3} S_i\psi_i \quad ,        \label{C8}
\end{equation}
where
\begin{equation}
   L_i=f_{ijk}x_j{\dot x}_k\quad ,
    \quad        S_i=-\fract{1}{2}if_{ijk}\psi_j\psi_k \quad .\label{C9}
\end{equation}
Furthermore,
\begin{eqnarray}
[L_i,x_j] &=& i f_{ijk}x_k\ ,\quad [S_i,\psi_j]= i f_{ijk}\psi_k\ ,\nonumber\\
{[}{\tilde Q}_3, x_i] &=& i f_{ijk}x_j\psi_k\ , \quad \{ {\tilde Q}_3,\psi_i\}=
L_i+2S_i                           \ .                          \label{ncon}
\end{eqnarray}
Both $L_i$ and $S_i$ are representations of the Lie algebra
${\cal G}$ since they obey
\begin{equation}
    [L_i,L_j]=if_{ijk}L_k\quad ,\quad    [S_i,S_j]=if_{ijk}S_k\ .   \label{C10}
\end{equation}

We note that the ${\tilde Q}_3$ supersymmetry does not close on the 
Hamiltonian, but instead we have
\begin{equation}
    \{ {\tilde Q}_3,{\tilde Q}_3\}={\vec J}^2+\fract{1}{3}
 {\vec S}^2 \quad , \quad J_i=L_i +S_i \quad , \label{C11}
\end{equation} where $J_i$ is the conserved charge associated with the 
$G$-invariance of the action (\ref{B1}).
The result here parallels the result of \cite{De.Ma.Pe.va}
for particle motion in the background field of a Dirac monopole.

We note  in passing also that ${\tilde Q}_3$ looks similar to the 
Kostant fermionic operator $K$ \cite{BR,K} 
\begin{equation}
   K=\rho_i\gamma^i-
\frac{i}{3! 2}f_{ijk}\gamma^i\gamma^j\gamma^k\ ,   \label{Kos} \end{equation}
where $\rho$ refers to some representation of ${\cal G}$ ({\it cf} $L_i$ in
(\ref{C9})\, ).
In (\ref{Kos}) the quantised fermion operators 
$\psi_i$ have been represented, $\psi_i \mapsto \frac{1}{\sqrt{2}}\gamma_i$, 
by the Dirac matrices of a euclidean space of dimension 
${\rm dim}\,{\cal G}$ which obey $\{\gamma_i,\gamma_j\}=2\delta_{ij}$.
However, (\ref{Kos}) implies $K^2=\vec{\rho}^2 + \fract{1}{3} \vec{S}^2$. 
Clearly $K$ 
is not proportional the supercharge ${\tilde Q}_3$ of (\ref{C1}), for any
representation $\rho$ of  $\cal G$.

\subsection{Other hidden supersymmetries}

We generalise now the previous paragraph to supercharges involving 
the higher order cocycles of the Lie algebra $\cal G$.
Instead of (\ref{C1}), we consider ${\tilde Q}_5$ as in (\ref{A3})
\begin{equation}
    {\tilde Q}_5=p_i f_{ijkp}\psi_{jkp}-\fract{1}{5}i\Omega_{ijkpq}\psi_{ijkpq}
                                              \quad ,             \label{D1}
\end{equation}
where $f_{ijkp}=f_{ijkp}(x)$,
\begin{equation}
   f_{ijkp}=f_{i[jkp]}       \ ,                                   \label{D2}
\end{equation}
we have used the abbreviation $\psi_{ijk\dots}=\psi_i\psi_j\psi_k\dots$ ,
and $\Omega_{ijkpq}$ is by definition totally antisymmetric in all its five 
indices. Instead of (\ref{D2}) we might consider the replacement of 
$\Omega_{ijkpq}$ by
\begin{equation}
    f_{[ijkp,q]}\ ,                                                  \label{D3}
\end{equation}
where $f_{ijkp,q}=\partial f_{ijkp}/\partial x_q$.
We now demand that ${\tilde Q}_5$ anticommutes with $Q=p_i\psi_i$ 
\begin{equation}
     \{ Q,{\tilde Q}_5\}=0                                         \label{D5}
\end{equation}
quantum mechanically for variety, and also because promotion of the classical 
calculation in this context is not in this instance trivial. Thus
\begin{eqnarray}
     \{ Q,{\tilde Q}_5\} &=& p_l\{ \psi_l,{\tilde Q}_5\}+
            [{\tilde Q}_5,p_l]\psi_l\nonumber\\
     &=& p_l(3p_if_{ijkl}\psi_{jk}-i\Omega_{ljkpq}\psi_{jkpq})
     +ip_l f_{ljkp,q}\psi_{jkpq} \ .                              \label{D7}
\end{eqnarray}
The first term of (\ref{D7}) can be eliminated by requiring that $f_{ijkl}$ is 
antisymmetric in $i$ and $l$, and, hence, using (\ref{D3}) that
\begin{equation}
     f_{ijkp}=f_{[ijkp]}\ .                                       \label{D8}
\end{equation}
The other result needed to secure (\ref{D5}) is the vanishing of the part of
\begin{equation}
     f_{ljkp,q}-\Omega_{ljkpq}                                      \label{D9}
\end{equation}
antisymmetric in $jkpq$. If we try the ansatz
\begin{equation}
    f_{ljkp}=\Omega_{ljkpq}x_q          \quad ,                  \label{D10}
\end{equation}
then this requirement is satisfied. Then
\begin{equation}
     {\tilde Q}_5=p_i\Omega_{ijkpq}x_q\psi_{jkp}-\fract{1}{5}
     i\Omega_{ijkpq}\Psi_{ijkpq}                                  \label{D11}
\end{equation}
is a hidden, $AdG$-invariant
 supercharge. This situation is one that applies to ${\tilde Q}_7$, 
involving $\Omega_7$, etc. Actually, since the charges are constructed using 
the structure constants and the higher-order
cocycles, there will be a hidden supercharge 
for each cocycle in a model with an arbitrary simple
compact group. We may then conclude the following: 

{\it The $Ad(G)$-invariant free supersymmetric particle model 
described by ${\cal L}=\frac{1}{2}\Phi_iD\Phi_i$ admits $l$ hidden 
supercharges ${\tilde Q}_s$, $s=1,\dots,l$. These are determined by 
the $l$ Lie algebra cohomology cocycles of the simple compact 
algebra $\cal G$ of rank $l$}.

If we had written $f_{[ijkp,q]}$ in the second term of (\ref{D1}), then 
(\ref{D9}) corresponds to the vanishing of the part of 
\begin{equation}
   f_{l[jkp],q}=f_{[ljkp,q]}                                     \label{D12}
\end{equation}
antisymmetric in $jkpq$. To extract the minimal condition, we write
\begin{equation}
    f_{[ljkp,q]}=\fract{4}{5} f_{l[jkp,q]}+\fract{1}{5}f_{[jkpq],l}\ .
\label{D13}
\end{equation}
Then (\ref{D12}) is equivalent to
\begin{equation}
    f_{[ljkp,q]}=f_{[jkpq],l}   \quad ,                       \label{D14}
\end{equation}
which is an analogue of (\ref{C6}) and, together with the complete 
antisymmetry of $f_{ijkl}$, means that $f_{ijkl}$ is a Killing-Yano tensor of 
valence four. This is consistent with Tanimoto's analysis
\cite{Ta}, although in a slightly 
different context. The previous result may be rephrased in the 
following form:

{\it The additional supersymmetry exists because each 
Lie algebra cocycle of order $2m-1$ provides a Killing-Yano tensor of valence 
$2(m-1)$ by the analogous to 
(\ref{D10}) since then}
\begin{equation}
   f_{ii_1\dots i_{2m-3},q}+f_{qi_1\dots i_{2m-3},i}=0\quad .                       
                                                             \label{4y}
\end{equation}

 One might look for different solutions, 
but we did not find any. For example, the condition (\ref{D14}) defeats the 
otherwise interesting ansatz $f_{ijkl}=f_{[ij}f_{kl]}$, where $f_{ij}$ is a 
second order Killing-Yano tensor. In fact, the Killing-Yano condition appears 
to be very restrictive, and it is likely that (\ref{C7}) and (\ref{D10}) 
are the only possible solutions for the cases considered.

\section{${\tilde Q}_3$ in the presence of a background field}

We investigate now to what extent it is possible to reproduce the 
analysis of sec. 3 in the $A\neq 0$ case.
When $Q$ is given by (\ref{B9}) and ${\tilde Q}_3$ by (\ref{C1}), where now 
$p_i={\dot x}_i+qA_i$ as in Sec. 2, we find classically that $\{ 
Q,{\tilde Q}_3\}=0$ requires 
\begin{eqnarray}
\{Q,{\tilde Q}_3 \} &=& -\psi_l{\dot x}_i f_{ij,l}\psi_j-
       i{\dot x}_l({\dot x}_i f_{il}-if_{ljk}\psi_j\psi_k)\nonumber\\
          & &+qf_{lj}F_{il}\psi_i\psi_j=0\quad ,
                                                               \label{E0}
\end{eqnarray}
so we have to impose (\ref{C4}), (\ref{C6}) and
\begin{equation}
      f_{l[j}F_{i]l}=0                               \label{E1}
\end{equation} 
where $F_{ij}$ is given by (\ref{efe}).
It is sensible to try first the solution
\begin{equation}
   f_{ij}=f_{ijk}x_k                                      \label{E3}
\end{equation}
of (\ref{C4}) and (\ref{C6}).
To satisfy (\ref{E1}) we may choose
\begin{equation}
F_{ij}= x_iy_j-x_jy_i    \quad ,                        \label{E4}
\end{equation}
where $y_i=d_{ijk}x_jx_k$, 
whenever there exists an invariant symmetric
tensor $d_{ijk}$ on the algebra (which excludes 
$su(2)$). This is true because of the 
identities $f_{ijk}x_jx_k=0=f_{ijk}x_jy_k$, the last one due to the 
invariance of the symmetric tensor $d_{ijk}$. However consistency 
with (\ref{efe}) demands that 
any ansatz for $F_{ij}$, and (\ref{E4}) in particular, obeys
\begin{equation}
       \partial_{[i}F_{jk]}=0   \quad .                           \label{E5}
\end{equation}

  The simple ansatz $F_{ij}= a(Z) f_{ijk}x_k$ where $Z=x_k x_k$,
satisfies (\ref{E5}) when the algebra is $su(2)$, in which case 
$a(Z) \propto Z^{-3/2}$, as is expected for the monopole \cite{De.Ma.Pe.va}.
But for $su(n)$, {\it e.g.} for $su(3)$, for which $Z=x_k x_k$ and
$Y=d_{ijk} x_i x_j x_k = x_k y_k$, the ansatz $F_{ij}= a(Z,Y) f_{ijk}x_k$
fails. We can see this by contracting
\begin{equation}
      0=\partial_{[i} F_{jk]}=2\frac{\partial \, a}{\partial \,  Z}
       x_{[i}f_{jk]l}x_l+ 
       3 \frac{\partial \, a}{\partial Y} y_{[i} f_{jk]l} x_l
       +a \, f_{ijk} \  , \label{EA}
\end{equation} with $y_i$, which gives
\begin{equation}
      \fract{2}{3} Y \, \frac{\partial \, a}{\partial \,  Z}
       f_{jkl}x_l+ y^2 \frac{\partial \, a}{\partial \, Y} f_{jkl}x_l 
       + a \, f_{jkl} y_l =0 \quad . \label{EB}
\end{equation} The independence of the tensors $f_{ijk}x_l$ and $f_{jkl}y_l$
now gives $a=0$. The method used here can be extended to show the failure also
of the more general ansatz
\begin{equation}
       F_{ij}= a(Z,Y) f_{ijk}x_k+ b(Z,Y) f_{ijk}y_k  \ . \label{EC}
\end{equation} The case of $su(2)$ of course escapes such a failure  because
$y_i$ then does not exist.

The choice (\ref{E4}) 
can be used for $su(n)$, $n\geq 3$:
it  obeys (\ref{E5}) directly but also allows us to write 
down suitable choices of $A_i$ for which (\ref{efe}) reproduces 
(\ref{E4}). For example, we may use 
\begin{equation}
   A_i=-\fract{1}{3}Yx_i          \quad .                     \label{E6}
\end{equation} 
More generally, any $A_i$ of the form
\begin{equation}
   A_i=\alpha(Z,Y)x_i+\beta(Z,Y)y_i                      \label{E7}
\end{equation}
gives $F_{ij}$ in suitable form:
\begin{equation}
F_{ij}=\left( 2\frac{\partial\beta}{\partial Z}-3
      \frac{\partial\alpha}{\partial Y}\right)(x_i y_j-x_jy_i)\ .
                                                             \label{E8}
\end{equation}
Thus, {\it for any $G$-invariant model (\ref{B1}) there exists an additional 
supersymmetry ${\tilde Q}_3$ given by}
\begin{equation}
   {\tilde Q}_3= p_if_{ijk}\psi_jx_k-\fract{1}{3}if_{ijk}\psi_i\psi_j\psi_k\ ,
\end{equation}
{\it determined by the structure constants $f_{ijk}$ of $\cal G$}.
The contribution proportional to $A_i$ disappears from the first term 
of (\ref{E8}) again because of the identities $f_{ijk}x_jx_k=0=f_{ijk}x_jy_k$.
But this does not mean that we have recovered the free case, because
$A_i$ is present in $Q=(p_i-qA_i)\psi_i$. ${\tilde Q}$ satisfies the relations
\begin{equation}
[{\tilde Q}_3,x_i]=if_{ijk}x_j\psi_k\ ,\;   
\{{\tilde Q}_3,\psi_i\}= L_i+2S_i\ ,\; 
\{ {\tilde Q}_3,{\tilde Q}_3\}=({\vec L}+{\vec S})^2+\fract{1}{3}{\vec S}^2
\, ,                                                   \label{intcom}
\end{equation}
where now $L_i=f_{ijk}x_jp_k$.

We have not found, however, an analogue of ${\tilde Q}_5$ for $q\neq 0$. If the 
Killing-Yano tensor is $f_{ijkl}=\Omega_{ijklm}x_m$ and $F_{ij}$ is 
proportional to (\ref{E4}), the condition corresponding to (\ref{E1}),
\begin{equation}
f_{l[jmn}F_{i]l}=0\ ,
\end{equation}
is not satisfied because, in contrast with 
$f_{ijk}x_j y_k=0$,  $\Omega_{ijklm}x_l y_m\neq 0$. It is possible that
the use of more general background fiels (see \cite{JWvH}) opens the
way to richer possibilities.

\section{$N=1$ Fermion superfields}

\subsection{Basic formalism}

We turn now to a different supersymmetric model in which hidden 
supersymmetries related to higher order cocycles also occur.
This model is a theory of fermions with all states in one energy level, and 
without bosonic dynamical variables \cite{Macfarlane}. 
Consider the $Ad(SU(n))$-invariant superspace 
Lagrangian given by
\begin{equation}
    {\cal L}=\fract{1}{2}\Lambda_i D\Lambda_i+\fract{1}{3!}igf_{ijk}
   \Lambda_i\Lambda_j\Lambda_k              \quad ,              \label{F1}
\end{equation}
where the $\Lambda_i=\Lambda_i(t,\theta)$ are $i=1,\dots,
{\rm dim}\, {\cal G}$ fermionic superfields,
\begin{equation}
       \Lambda_i(t,\theta) = \psi_i(t)+\theta B_i(t)\ ,\quad
       D\Lambda_i = B_i-i\theta {\dot \psi}_i       \ .        \label{F2}
\end{equation}
The expansion of (\ref{F1}) may be written as ${\cal L}=K+\theta L$,
 where now
\begin{eqnarray}
      K &=& \fract{1}{2}\psi_i B_i+\fract{1}{3!}ig\psi_i\psi_j\psi_k
           \quad ,     \nonumber\\
     L &=& \fract{1}{2}(i{\dot\psi}_i\psi_i+B_i B_i+ig f_{ijk}\psi_i\psi_j
         B_k)   \ .                                                \label{F3}
\end{eqnarray}
As in Sec. 2, we may use the expression of $K$ and $L$ in 
Noether's theorem (see (\ref{B8})) for the variations
\begin{equation}
\delta\psi_i=-\epsilon B_i\ ,\quad \delta B_i=i\epsilon{\dot\psi}_i
\quad  ,
\label{F4}
\end{equation}
to obtain
\begin{equation}
   Q=\fract{1}{6}igf_{ijk}\psi_i\psi_j\psi_k      \ .      \label{F5}
\end{equation}
All the bosonic components $B_i$ are auxiliary. Using their 
Euler-Lagrange equations to solve for $B_i$,
 one obtains the classical Lagrangian 
$L_c=\frac{1}{2} i \psi_i {\dot \psi}_i$.
The classical Hamiltonian vanishes identically \footnote{There is an 
$S_iS_i$ part,  which is proportional to $f_{mij}
f_{mkl}\psi_i\psi_j\psi_k\psi_l$ and which is zero classically by virtue of 
the Jacobi identity and the fermionic character of the $\psi$'s.}, but the 
quantum Hamiltonian, defined by $H=Q^2$, and computed using $\{ \psi_i,
\psi_j\}=\delta_{ij}$, is not zero but a constant, 
\begin{equation}
       Q^2=\fract{1}{48}\, g^2 \, n(n^2-1)                       \label{F6}
\end{equation}
for $su(n)$.

\subsection{Hidden supersymmetries in the fermionic model}

As for the (\ref{B1}) model for $q=0$, there exist in this case
additional supercharges $\tilde Q$ for every non-trivial cocycle of any $su(n)$
 Lie algebra. To see this, let $\Omega^{(2m-1)}_{i_1\dots i_{2m-1}}$ be a 
non-trivial cocycle corresponding to an invariant symmetric
tensor $t$ of order $m$. If $t$ has components
$t_{l_1\dots l_m}$, then the $(2m-1)$-order cocycle is given by\footnote{For 
the relation among symmetric tensors and cocycles of generic compact simple 
Lie algebra $\cal 
G$ see \cite{tensors}, \cite{compilation} and \cite{casimirs}.}
\begin{equation}
      \Omega^{(2m-1)}_{i_1\dots i_{2m-1}}=f^{l_i}_{[i_1i_2}\dots
    f^{l_{m-1}}_{i_{2m-3}i_{2m-2}} t_{i_{2m-1}]l_i\dots l_{m-1}}
                                   \ .                   \label{F6A}
\end{equation}
 Using it, we form
\begin{equation}
  {\tilde Q}_{2m-1}=\Omega^{(2m-1)}_{i_1\dots i_{2m-1}}\psi_{i_1\dots
  i_{2m-1}}                           \ ,                         \label{G1}
\end{equation}
where 
\begin{equation}
   \psi_{i_1\dots i_{2m-1}}\equiv \psi_{i_1}\dots \psi_{i_{2m-1}}\ .\label{G1A}
\end{equation}
We compute the quantum anticommutator of $Q$ in (\ref{F5}) and 
${\tilde Q}_{2m-1}$ by expressing the products $\psi_{i_ii_2i_3}
\psi_{j_1\dots j_{2m-1}}$ and $\psi_{j_1\dots j_{2m-1}}\psi_{i_ii_2i_3}$ as 
linear combinations of completely antisymmetrised products of $\psi$'s. This 
can be done by repeated use of the identities (deduced from $\{ \psi_i 
,\psi_j\}=\delta_{ij}$)
\begin{eqnarray}
     \psi_i\psi_{j_1\dots j_p} &=& \psi_{ij_1\dots j_p}+\fract{1}{2}p
      \delta_{[ij_1}\psi_{j_2\dots j_p]}\ ,\nonumber\\
    \psi_{j_1\dots j_p}\psi_i &=& \psi_{j_1\dots j_pi} +\fract{1}{2}p
      \psi_{[j_1\dots j_{p-1}}\delta_{j_p]i}\ .
\end{eqnarray}
Then we easily find that the terms with no $\delta$'s or an even number of 
them vanish identically because the two contributions coming from the 
anticommutator cancel each other. So we are left with
\begin{eqnarray}
        \{ Q,{\tilde Q}_{2m-1} \} &=& \fract{1}{6}ig[3(2m-1)f_{ki_2i_3}
    \Omega^{(2m-1)}_{kj_2\dots j_{2m-1}}\psi_{i_2i_3j_2\dots j_{2m-1}}
           \nonumber \\
     & & -\fract{1}{4}\frac{(2m-1)!}{(2m-4)!}f_{klm}
    \Omega^{(2m-1)}_{klmj_4\dots j_{2m-1}}\psi_{j_4\dots j_{2m-1}}]
  \ .                                                             \label{G2}
\end{eqnarray}
The first term in (\ref{G2}) vanishes due to the Jacobi identity
(since the indices $i_2i_3j_2\dots j_{2m-1}$ are antisymmetrised
due to the presence of the $\psi$'s \, )
and the second also vanishes because the maximal contraction of indices 
among the above two  $su(n)$ cocycles of different order gives 
zero \cite{compilation}. Hence, 
{\it the $l$ ${\tilde Q}_{2m-1}$ define new conserved fermionic 
charges of higher order}. As in the case of $Q$ in (\ref{F5}), 
they square to a constant. For example, let us consider the 
case of ${\tilde Q}_5$ for $su(n)$, $n\geq 3$. The square 
of ${\tilde Q}_5$ is given by
\begin{equation}
  {\tilde Q}_5^2=\Omega^5_{i_1\dots i_5}\Omega^5_{j_1\dots j_5}
       \psi_{i_1\dots i_5}\psi_{j_1\dots j_5}\ .            \label{G3}
\end{equation}
It is shown in \cite{Az.Ma} (see also \cite{compilation}) that 
this square is a number. Hence, $Q$, ${\tilde Q}_5$ and ${\bf I}$ 
close into a superalgebra.

\section{$N=2$ Fermion superfields}

\subsection{Basic formalism}

We now consider a purely fermionic model with two standard supersymmetries
\cite{Macfarlane} (see also \cite{CAMP}). The supersymmetry algebra 
in terms of the covariant derivatives for this model is
\begin{equation}
   D = \partial_\theta-i\theta^*\partial_t\ ,\quad 
 D^*=-\partial_{\theta^*}+i\theta\partial_t           \ ,\quad 
\{ D,D^*\} =  2i\partial_t\ .                               \label{6.0}
\end{equation} 
The $N=1$ superfields $\Lambda_i$ are replaced by $N=2$ 
superfields 
$\Psi_i=\Psi_i(t,\theta,\theta^*)$, $i=1,\dots,{\rm dim}\, {\cal G}$,
to which the chirality condition $D^*\Psi_i=0$ is imposed. This of course means that $\Psi^*_i$ obeys 
$D\Psi^*_i=0$ and is antichiral. Solving as usual the chirality condition we 
obtain the superfield expansions
\begin{eqnarray}
      \Psi_i &=& e^{i\theta^*\theta\partial_t}(\mu_i-\theta B_i)=
             \mu_i-\theta B_i+i\theta^*\theta {\dot\mu}_i \nonumber\\
       \Psi^*_i &=& \mu^*_i-\theta^* B^*_i-i\theta^*\theta{\dot\mu}^*_i \ ,
                                                                   \label{6.1}
\end{eqnarray}
where $\mu_i$ are fermionic and $B_i$ are bosonic. The following 
$Ad(G)$-invariant superspace action has the property that the $B$'s are 
non-dynamical, and includes an interaction term:
\begin{eqnarray}
    S &=& -\fract{1}{2}\int dt d\theta d\theta^*\Psi_i\Psi_i^* \nonumber\\
   & & +\fract{1}{6}\int dt d\theta iC_{ijk}\Psi_i\Psi_j\Psi_k
          +\fract{1}{6}\int dt d\theta^* iC_{ijk}\Psi_i^*\Psi_j^*\Psi_k^* \ ,
\label{6.2}
\end{eqnarray}
where $C_{ijk}$ are the structure constants of the Lie algebra $\cal G$. The 
$Ad(G)$-invariant component Lagrangian is given by
\begin{equation}
      L=\fract{1}{2}i(\mu_i^*{\dot\mu}_i+\mu_i{\dot\mu}_i^*)+ 
\fract{1}{2}B^*_iB_i-\fract{1}{2}iC_{ijk}\mu_i\mu_j B_k-\fract{1}{2}iC_{ijk}
  \mu_i^*\mu_j^* B_k^*\ .\label{6.3}
\end{equation}
The Euler Lagrange equations of the $B_i$, $B_i^*$ are algebraic and can be 
used to eliminate these variables. Then, by writing 
$J_i=-\fract{1}{2}iC_{ijk}\mu_j\mu_k$, the Lagrangian becomes
\begin{equation}
    L=\fract{1}{2}i(\mu_i^*{\dot\mu}_i+\mu_i{\dot\mu}_i^*)-2J_iJ_i^* \ .
      \label{6.4}
\end{equation}
The canonical formalism yields the Dirac brackets 
\begin{equation}
       \{\mu_i,\mu^*_j\}=-i\delta_{ij},\quad \{\mu_i,\mu_j\}=0,\quad 
     \{\mu_i^*,\mu_j^*\}=0\ , \label{6.5}
\end{equation}
and classically we have 
\begin{equation}
    H=2J_iJ_i^* \ .                 \label{6.6}
\end{equation}
The non-zero supersymmetry variations of the fields $\mu_i$ and $B_i$ are 
given by
\begin{eqnarray}
    \delta_{\epsilon^*}B_i&=&-2i\epsilon^*{\dot\mu}_i\ ,\quad \delta_{\epsilon^*}
     \mu^*_i=\epsilon^*B^*_i\ ,\nonumber\\
  \delta_{\epsilon}B_i^*&=&-2i\epsilon{\dot\mu}^*_i\ ,\quad \delta_{\epsilon}
     \mu_i=\epsilon B_i \ .          \label{6.7}
\end{eqnarray}
These variations correspond, via Noether's theorem, to the conserved charges
\begin{equation}
     Q=\fract{1}{3}iC_{ijk}\mu_i\mu_j\mu_k\ ,\quad 
Q^*=\fract{1}{3}iC_{ijk}\mu_i^*\mu_j^*\mu_k^*\ .       \label{6.8}
\end{equation}

When we quantise the theory we shall regard  the $\mu_i$ as the creation 
operators. Hence, to avoid 
confusion, the following replacements will be made from now on: 
$\mu^*_i=\pi_i$, 
$\mu_i=c_i$, as in \cite{CAMP}, so that $\pi^*=c_i$.
Thus in the quantum theory, we have the anticommutation relations
\begin{equation}
    \{ c_i \; , \; \pi_j \}=\delta_{ij}\ .          \label{6.9A}
\end{equation}
Also the quantum mechanical Hamiltonian is defined via
\begin{equation}
    \{ Q \; , \; Q^* \}=2H_q \ ,          \label{6.10A}
\end{equation}
which gives the result 
\begin{equation}
   H_q= \{ J_i \; , \; J_i^* \}- \fract{1}{6} c^2 \; , 
               \label{6.11A} 
\end{equation} where $c^2=C_{ijk} C_{ijk}$ ($=n(n^2-1)$ for $su(n)$), 
and where it should be noted that $J_i$ and $J_i^*$ do not commute. 
One might expect that that 
$H_q$ is closely related to the quadratic 
Casimir operator $X^2= X_i X_i$ of ${\cal G}$, where  
$X_i=-iC_{ijk}\, c_j \, \pi_k\ $. It is simple to 
confirm this for the case of $su(n)$ by proving the following identities
\begin{eqnarray}
  X_i X_i & = & nN -2J_i J_i^* \nonumber \\
          & = & n(n^2-1-N) -2J_i^* J_i \quad , \label{6.12A} 
\end{eqnarray} 
where $N=c_i \pi_i$ is the total fermion number operator, and
$\pi_i c_i=(n^2-1-N)$. These allow the commutator and anticommutator of 
$J_i$ and $J_i^*$ to be calculated, and give rise to the result
\begin{equation}
   H_q =\fract{1}{3} \, n(n^2-1) - X_i X_i \ .
\label{6.13A} 
\end{equation} 
The results (\ref{6.11A}) and (\ref{6.13A}), viewed together,
seem a strangely related pair. However their agreement, as well 
as the correctness of (\ref{6.12A}), can easily be confirmed by 
considering actions of the operators in question on each of the 
fermion number $N=0,\; 1,\; 2,\; 3$ states of the simple but 
non-trivial $SU(2)$ version of the theory. Further, having set 
out from the definition (\ref{6.10A}) of $H_q$, we know that all 
energy eigenvalues are non-negative. 

In addition to 
\begin{equation}
 q_{30}=Q/2 = \fract{1}{3!}i C_{ijk} c_i c_j c_k \quad , \quad 
q_{03}=q_{30}^* =Q^*/2 \ ,
\label{6.14A} 
\end{equation} 
in which the first and second subscripts indicate the numbers of
$c_i$ and $\pi_i$ factors respectively, two further fermionic operators occur
naturally:
\begin{equation}
     q_{21}=\fract{1}{2}\, i C_{ijk}c_ic_j\pi_k\ ,\quad 
q_{12}=\fract{1}{2}\, i C_{ijk}c_i\pi_j\pi_k\ .                 \label{6.15A}
\end{equation}
These operators each anticommute with each of $q_{30}$ and $q_{03}$ and obey
\begin{equation}
\{  q_{21} \; , \; q_{12} \} = \fract{1}{2} X_i X_i \quad , \quad q_{21}^2=0
\quad , \quad q_{12}^2=0 \ .
\label{6.16A} \end{equation}
It follows that $ q_{21}$ and $ q_{12}$ commute with $X_i X_i$ 
and with $H$.

We have found a second supersymmetry 
which anticommutes with the original one; its closure does not 
give a new operator independent of $H$. However, in view of the 
results (\ref{6.11A}) and (\ref{6.13A}), it is not appropriate 
to say that our theory has $N=4$ supersymmetry. We have simply found 
two additional supercharges naturally associated with the structure 
constants of ${\cal G}$, a consideration that is built on significantly 
in the next subsection.

\subsection{Hidden supersymmetries}

One obvious question asks whether it is possible to find new 
supercharges that generalize those of Sec. 6.1. Consider the case of 
charges constructed using the five-cocycle $\Omega_{i_1\dots i_5}$ 
rather than the three-cocycle $C_{ijk}$. 
An analysis of the possibilities, $q_{50}$, $q_{41}$, $q_{32}$, $q_{23}$, 
$q_{14}$, $q_{05}$, leads one to conclude that only
\begin{equation}
 q_{05}=\fract{1}{5!}i\Omega_{i_1\dots i_5}\pi_{i_1}\dots \pi_{i_5}\ , \quad
 q_{50}=\fract{1}{5!}i\Omega_{i_1\dots i_5}c_{i_1}\dots c_{i_5}  \label{6.14}
\end{equation}
are hidden conserved  supercharges because only they
anticommute with $q_{21}$ and $q_{12}$.

Moreover, we have the following general result:

{\it The $l={\rm rank}\,{\cal G}$ pairs of fermionic charges
\begin{eqnarray}
     q_{0,2m-1}&=&\frac{i}{(2m-1)!}\Omega_{i_1\dots i_{2m-1}}
          \pi_{i_1}\dots \pi_{i_{2m-1}}\ ,\nonumber \\
    q_{2m-1,0}&=&\frac{i}{(2m-1)!}\Omega_{i_1\dots i_{2m-1}}
          c_{i_1}\dots c_{i_{2m-1}}   \ ,             \label{6.15}
\end{eqnarray}
determined by the $(2m-1)$-cocycles of the Lie algebra $\cal G$, where the 
allowed values of $m$ depend on the specific $\cal G$ considered, 
also commute with $q_{21}$ and $q_{12}$, and hence they commute with $H_q$ 
and are conserved supercharges}.

{\it Proof}: Let us restrict ourselves to $q_{2m-1,0}$ (the case $q_{0,2m-1}$ 
is completely analogous). Consider first
\begin{equation}
        \{ q_{2m-1,0},q_{12}\}\propto \Omega_{i[i_2\dots i_{2m-1}}C_{j_1 j_2]i}
           c_{j_1}c_{j_2}c_{i_2}\dots c_{i_{2m-1}}\ ,    \label{6.16}
\end{equation}
where the antisymmetrization is forced by the presence of the $c$'s. This 
expression vanishes by the $G$-invariance of $\Omega$, since this implies
\begin{equation}
          \Omega_{i[i_1\dots i_{2m-2}}C_{i_{2m-1}]ij}=0\ . \label{invariance}
\end{equation}
Hence,
$\{q_{2m-1,0},q_{12}\}=0$. Now we have to check that the following anticommutator 
also vanishes:
\begin{eqnarray}
       \{ q_{2m-1,0},q_{21}\}&\propto &\Omega_{ii_2\dots i_{2m-1}}C_{ij_1j_2}
      c_{j_1}c_{i_2}\dots c_{i_{2m-1}}\pi_{j_2}\nonumber\\
       & & +(m-1)\Omega_{ij i_3\dots i_{2m-1}}C_{ij j_1}c_{j_1}c_{i_3}\dots 
c_{i_{2m-1}}\ .                                      \label{6.17}
\end{eqnarray}
The first term vanishes due to the antisymmetry in the indices 
$j_1,i_2,\dots,i_{2m-1}$ and the invariance of $\Omega$. To show that the 
second term also vanishes, we have to prove that
\begin{equation}
           D\equiv \Omega_{iji_1\dots i_{2m-3}}C_{i_{2m-2}ij} c_{i_1}\dots 
c_{i_{2m-2}}                                            \label{6.18}
\end{equation}
is equal to zero. Indeed, the invariance of $\Omega$ allows us to write it in 
terms of the $C$'s and the invariant symmetric tensor
$t_{l_1\dots l_{m}}$ (see (\ref{F6A})) 
without having to involve $i,j$ in the antisymmetrization, so we have
\begin{equation}
    D= C_{ii_1l_1}C_{i_2i_3l_2}\dots C_{i_{2m-4} i_{2m-3}l_{m-1}} t_{l_1\dots 
l_{m-1} j}C_{i_{2m-2}ij}c_{i_1}\dots c_{i_{2m-2}}\ .           \label{6.19}
\end{equation} 
Now, using the Jacobi identity
\begin{equation}
C_{ii_1l_1}C_{i_{2m-2}ij}=C_{ii_{2m-2}l_1}C_{i_1ij}+C_{ii_1i_{2m-2}}C_{l_1ij}
                                    \quad ,      \label{6.20}
\end{equation}
we arrive at
\begin{eqnarray}
         D &=& C_{ii_{2m-2}l_1}C_{i_2i_3l_2}\dots C_{i_{2m-4}i_{2m-3}l_{m-1}} 
t_{l_1\dots l_{m-1}j}C_{i_1 ij}c_{i_1}\dots c_{i_{2m-2}}\nonumber\\
   & & +C_{ii_1i_{2m-2}}C_{i_2i_3l_2}\dots C_{i_{2m-4}i_{2m-3}l_{m-1}} 
t_{l_1\dots l_{m-1}j}C_{l_1 ij}c_{i_1}\dots c_{i_{2m-2}}\ .   \label{6.21}
\end{eqnarray}
The first term of this expression is equal to $-D$ due to the presence of 
$c_{i_1}$ and $c_{i_{2m-2}}$, and the second term vanishes because $C_{l_1ij}$ 
is antisymmetric in $l_1,j$ whereas $t_{l_1\dots l_m-1 j}$ is symmetric in these 
indices. So we have $D=-D$, $D=0$, $\{ q_{2m-1,0}\, ,q_{21}\}=0$ and 
$[q_{2m-1,0}\, ,H_q]=0$, {\it q.e.d.}.

Hence, the following result follows:

{\it For every simple Lie algebra $\cal G$ of rank $l$, 
the model (\ref{6.2}) has a series of $2l$ conserved supercharges 
that are constructed from the primitive cocycles of $\cal G$, 
which include the supersymmetry generators}.

\section{Hidden supersymmetries as Noether charges}

All the hidden supercharges discussed can be shown to be Noether charges 
associated with actual supersymmetries of the actions of the models in 
question. 
One way to realise  this is to use the quantum 
commutator of the supercharge and the variables of the model to extract the 
variations. Explicitly, if $\tilde Q$ is the conserved supercharge and 
$u$ is a generic component field in the model, its variation may be defined by 
\begin{equation}
          \delta_{\tilde \epsilon} u= [{\tilde \epsilon}{\tilde Q},u]\ ,   
\label{N1}
\end{equation}
where $\tilde \epsilon$ is the corresponding fermionic parameter. 

   If $\tilde Q$ is a symmetry of the classical action $S=\int dt L$, 
$\delta_{\tilde\epsilon}S=0$ for the constant parameter $\tilde \epsilon$.
This  means that, if we allow ${\tilde \epsilon}$ to become a function of $t$, 
its variation will be of the form
\begin{equation}
    \delta_{\tilde \epsilon} S=\int dt (-i{\dot {\tilde \epsilon}}{\tilde Q})
                          \label{N2} 
\end{equation}
ignoring boundary terms, where $\tilde Q$ is the conserved Noether charge for 
the symmetry (\ref{N1}). Indeed, from (\ref{N2}) we get 
$\delta_{\tilde\epsilon}S=\int dt (i{\tilde \epsilon}{\dot {\tilde Q}})$, and
since for 
solutions of the Lagrange equations $\delta_{\tilde\epsilon}S$ must be zero for
any $\delta$, it follows that $ {\dot {\tilde Q}}=0$ and $\tilde Q$ is the
conserved charge. This procedure is particularly suitable 
when, as here, the complications addressed in \cite{Ma.Mo} do not arise.

 We now give the variations obtained by using (\ref{N1}). In the 
bosonic case with $A_i\neq 0$, use of (\ref{B5}) yields
\begin{equation}
        \delta_{\tilde \epsilon} x_i=-i{\tilde \epsilon}_i 
      f_{ijk}x_k\psi_j\ ,\quad 
   \delta_{\tilde \epsilon} \psi_i={\tilde \epsilon} {\dot x}_j f_{jik}x_k
  -i{\tilde \epsilon} f_{jki}\psi_j\psi_k                          \label{N3}
\end{equation}
for the variation induced by ${\tilde Q}_3$ (eq. (\ref{C8})). If now we
put $\epsilon =\epsilon(t)$ and ignore boundary terms in 
the integrand, we do find $\delta L=-i{\dot \epsilon}{\tilde Q}_3$, 
recovering ${\tilde Q}_3$ as the Noether charge. 
The same applies to the other supercharges below.

Consider ${\tilde Q}_5$ (eq. (\ref{D1})). The corresponding 
formulae for the variations for the $A=0$ model of Sec. 3.2 are, 
\begin{equation}
\delta_{\tilde \epsilon} x_i=
-i{\tilde \epsilon}\Omega_{ijkpq}x_q\psi_{jkp}\ ,\quad
\delta_{\tilde \epsilon}\psi=
3{\tilde \epsilon} {\dot x}_j\Omega_{jkpiq}x_q\psi_{kp}-i{\tilde \epsilon}
\Omega_{jkpqi}\psi_{jkpq}\ .                                \label{N4}
\end{equation}
Note that in the above variations $\delta\psi_i$ involves 
the derivative of $\dot 
x$ (mathematically, this means that succesive tangent spaces 
-- jet spaces -- are needed to define the action of the 
${\tilde Q}$'s). This is not the case for the 
two purely fermionic models for which the 
canonical quantum commutators give
\begin{equation}
    \delta_{\tilde \epsilon}\psi_i=(2m-1){\tilde \epsilon}
\Omega_{i_1\dots i_{2m-2}i} \psi_{i_1\dots i_{2m-2}}    \label{N5}
\end{equation}
for ${\tilde Q}_{2m-1}$ (eq. (\ref{G1})), for the $N=1$ model 
of Sec. 5.1, and 
\begin{eqnarray}
\delta_{{\tilde \epsilon}^*} \mu_i &=& \frac{i{\tilde \epsilon}^*}{(2m-2)!}
\Omega_{i_1\dots i_{2m-2}i} \mu_{i_1}^*\dots \mu_{i_{2m-2}}^*\ ,\quad 
\delta_{{\tilde \epsilon}^*} 
\mu_i^*=0\ ,\nonumber \\
    \delta_{\tilde \epsilon}\mu_i^* &=& \frac{i{\tilde \epsilon}}{(2m-2)!}\Omega_{i_1\dots 
i_{2m-2}i}\mu_{i_1}\dots \mu_{i_{2m-2}}\ ,\quad \delta_{\tilde \epsilon}\mu_i=0 \label{N6}
\end{eqnarray}
for the variations produced, respectively, by the supercharges $q_{0,2m-1}$ and 
$q_{2m-1,0}$ (eqs. (\ref{6.15})), of the N=2 model of Sec. 6.1. 

   To proceed further, consider first the closure of
$\delta_\epsilon$, $\delta_{\tilde\epsilon}$ on $x_i$, say, for the $A\neq 0$ 
model in Sec. 4. First we find 
\begin{equation}
        [\delta_\epsilon,\delta_{\tilde\epsilon}]x_i=0  \label{N7}
\end{equation} 
reflecting the fact that $\{ Q,{\tilde Q}_3\}=0$ (eq. (\ref{E0})). For 
$[\delta_{{\tilde\epsilon}{'}},\delta_{\tilde\epsilon}]$ we find 
\begin{equation}
[\delta_{{\tilde\epsilon}{'}},\delta_{\tilde\epsilon}]x_i=-2i{\tilde\epsilon} 
{\tilde\epsilon}{'}f_{ijk}J_j x_k\ ,                      \label{N8} 
\end{equation}
where 
\begin{equation}
      J_i = f_{ijk} x_j {\dot x}_k - \fract{1}{2}if_{ijk}\psi_j\psi_k 
  \label{N9} 
\end{equation}
is the conserved charge ({\it cf}. (\ref{C11}) \, ) 
associated with the adjoint transformations $\delta 
x_i=f_{ijk}a_jx_k$, $\delta\psi_i=f_{ijk}a_j\psi_k$ which leave the Lagrangian 
(\ref{B4}) invariant. This of course agrees with the variations on $x_i$
induced by the operator
$ \{{\tilde Q}_3,{\tilde Q}_3\}$
expressed in the form ({\it cf}. again (\ref{C11}) \, )
${\vec J}^2 +\fract{1}{3}
{\vec S}^2$.                                 
A similar analysis can be performed for the corresponding actions 
upon $\psi_i$, for which we get 
\begin{equation}
[\delta_{{\tilde\epsilon}{'}},\delta_{\tilde\epsilon}]\psi_i=
-2i{\tilde\epsilon} 
{\tilde\epsilon}{'}f_{ijk}J_j \psi_k-2i{\tilde\epsilon}
{\tilde\epsilon}{'}f_{jlm}f_{jik}x_mx_k{\dot\psi}_l  \; ,
       \label{N11}
\end{equation}
in which we see the second term is zero on shell, using
${\dot \psi}_l=qF_{lj}\psi_j$
where $F_{lj}$ is given by (\ref{E4}).
 
\section{Concluding remarks}

We have shown in this paper that there exist `hidden' 
supersymmetric fermio\-nic charges associated with the Lie algebra cohomology 
cocycles of the symmetry group for three simple $Ad(G)$-invariant 
supersymmetric models, one with bosonic and fermionic coordinates and 
two (for $N=1$ and $N=2$) 
with only fermionic dynamical coordinates. 
For the first one, and in the free case, there are $l$ 
additional supersymmetries the existence of which is tied to the Killing-Yano 
tensors of valence $(2m-2)$ that may be constructed from the 
$(2m-1)$ Lie algebra cohomology cocycles. In the interacting  $A\neq 0$ 
case the same procedure seems to allow 
for only one additional supersymmetry, associated with the structure constants 
$f_{ijk}$ of the Lie algebra $\cal G$ considered. In the  $N=1$
fermionic model, 
at least in the case of ${\cal G}=su(n)$ 
$l$ additional symmetries may be constructed directly from 
its cocycles. In general, these purely fermionic ${\tilde Q}$'s 
depend only on the cohomology of $\cal G$ or, equivalently, 
on the topology of the corresponding compact group $G$. In this 
sense the additional supercharges may be traced to the topology of $G$;
however, they may be seen to generate continuous symmetries of the 
system and may be obtained from Noether's theorem. In the $N=2$ 
case it is shown that the standard supercharges are in fact part 
of a series of $2l$ conserved supercharges that can be 
constructed from the $l$ cocycles of $\cal G$. 

Summarising, our analysis shows that the hidden symmetries appearing in 
$Ad(G)$-invariant models are in fact supersymmetries
because they stem from the existence of the $G$-invariant odd 
skewsymmetric tensors associated with the cohomology of $\cal G$. 
In this sense, these additional fermionic charges could have been introduced 
directly, even before having a 
supersymmetric model, as in the (${\cal G}=su(2)$ \,)
analysis of \cite{Spec} of the results of 
\cite{De.Ma.Pe.va}. However, the fact that they square to a Casimir has more 
to do with the structure of the cocycles themselves \cite{compilation}, and 
hence with the structure of the generic symmetry group, than with any other 
considerations of $su(2)$. In fact, the expression 
(\ref{C11}) (see also (\ref{intcom})), that holds for {\it any} 
Lie algebra $\cal G$ explains why {\it e.g.} $\{ {\tilde Q}_3,{\tilde Q}_3\}$
is given by a Casimir. Also, our analysis in sec. 4 shows that the rotational 
symmetry of the model in \cite{De.Ma.Pe.va} does not play an essential role
(cf. \cite{Spec}), being just a result of the fact that the 
fully antisymmetric tensor in $3$-dimensions provides the structure 
constants.

\vskip 1.5cm
{\bf Acknowledgements}. This work was partly supported by the
DGICYT, Spain ($\#$PB 96-0756), the Junta de Castilla y Le\'on,
Spain ($\#$C02/199) and PPARC, UK.

\end{document}